\documentclass[12pt]{article}
\usepackage{graphicx}
  \newcommand{\sig}{ \mbox{\boldmath{$\sigma$}}}
   \newcommand{\bXi}{ \mbox{\boldmath{$\Xi$}}}
 
 \title{Suppression  of spin-orbit effects in 1D system
}
\author{M.V. Entin
and  L.I. Magarill \\\it Institute of Semiconductor Physics,\\
\it Siberian Branch of Russian Academy of Sciences\\\it - 13
Lavrent'eva, Novosibirsk, 630090, Russia}

\begin{document}
\maketitle PACS: 71.70.Ej;  73.63.-b; 73.63.Nm\\ \\ Submitted to
Europhysics Letters
\begin{abstract}
We report the absence of spin effects such as spin-galvanic
effect, spin polarization and spin current under static electric
field and inter-spin-subband absorption in 1D system with
spin-orbit interaction of arbitrary form.   It was also shown that
the accounting for the direct interaction of electron spin with
magnetic field violates this statement.
\end{abstract}
The spin-orbit (SO) interaction in a 2D system underlies various
spin control methods owing to the coupling between translational
and spin degrees of freedom. Such effects have been studied as
 spin-galvanic effect \cite{gan1}-\cite{gan3}, spin
polarization \cite{edel1}-\cite{chap} and spin current
\cite{schliem} under static electric field, spin polarization
under action of electromagnetic wave \cite{edel2}. The one
dimensional system seems to be more suitable for this purpose due
to more strong correlation between the spin and the wire
direction. This stimulates to examine the similar problems in 1D
systems.

We consider the 1D Hamiltonian
\begin{equation}\label{1}
       {\cal H}=\frac{p^2}{2m}+V(x)+{\cal H}_{SO} \end{equation}
with the most general form of SO interaction
       \begin{equation}
       \label{1'} {\cal H}_{SO}=\{({\bf a}(x)\mbox{\boldmath{$\sigma$}}),p\},
\end{equation}
where $\mbox{\boldmath{$\sigma$}}$ are the Pauli matrices, the
figure brackets denote the symmetrization procedure, vector ${\bf
a}(x)$ is an arbitrary function of coordinate $x$ along the wire.
The Hamiltonian (\ref{1'}) originates from different approaches
related with SO interaction  in 1D systems.  In general, it does
not conserve the spin and hence one can expect the above mentioned
effects in the frameworks of this Hamiltonian. However, we have
found that in a strictly 1D system with the SO Hamiltonian
(\ref{1'}) these effects vanish.

\section*{One-dimensional systems obeying the Hamiltonian (\ref{1}-\ref{1'})}
For example, let us consider the 1D quantization of
 the 2D Rashba Hamiltonian \cite{rash1},\cite{rash2}
  \begin{equation} \label{RH}
\hat{\cal H}_{SO} =
  \alpha_R(\mbox{\boldmath{$\sigma$}}[{\bf p}\times{\bf
  n}]),
\end{equation}
where ${\bf p}$ is the 2D electron momentum, and ${\bf n}$ is the
normal to the plane of the system (axis $z$). The size
quantization  in  $y$ direction leads to the reduction of the
Hamiltonian (\ref{RH}) for the lowest subband:
  \begin{equation} \label{RH1}
\hat{\cal H}_{SO} = -
  \alpha_R\sigma_yp_x,
\end{equation}
where $p_x=p$ is the 1D momentum. In this case the vector ${\bf
a}=(0,-\alpha_R,0)$.

The similar Hamiltonian arises in cubic crystals with no inversion
symmetry from the spin-orbit
  term in the bulk
Hamiltonian of Dresselhaus \cite{dres}
$$\hat{\cal H}_{SO}^{(b)}=(\delta_0/2)\varepsilon_{ijk}\sigma_i\lambda_{rsk}p_rp_jp_s,$$
where $p_i$ is 3D electron momentum. The symmetric  in all indexes
tensor $\lambda_{rsk}$ characterizes the anisotropy of the
crystal. In the principal crystal axes $\lambda_{rsk}$ has the
only non-zero component $\lambda_{123}=1$, in the general case
$\lambda_{ijk} =\sum_{n\neq m\neq l} W_{in} W_{jm} W_{kl}$, where
$W_{ij}$ is the rotation matrix from the frame of reference of
  crystallographic axes to the laboratory system. The confinement along two
directions, say $y$ and $z$, converts   $\hat{H}_{SO}^{(b)}$ into
the 1D  Hamiltonian, linear in the momentum $p$. The form of this
Hamiltonian depends on the orientation of the wire  relative to
the principal crystal axes.

 As the result of quantization we obtain
 \begin{eqnarray}\label{1DDres}
    {\bf
    a}=\delta_0\Big(2\lambda_{xyz}(\overline{p_y^2}-\overline{p_z^2}),~~
      2\lambda_{xxz}\overline{p_z^2}-
      \lambda_{zzz}\overline{p_z^2}-\lambda_{yyz}\overline{p_y^2},~~
  \lambda_{yyy}\overline{p_y^2}- 2
   \lambda_{yxx}\overline{p_y^2}+\lambda_{zzy}\overline{p_z^2}\Big),
\end{eqnarray}
where the overline means the averaging with the wave function of
the ground state in quantum well.

  In the considered cases the vector $\bf a$ is constant.
  More general is the situation of  curved  wire, in which the vector ${\bf a}$
  becomes variable in accordance with  the change of local direction of the axis $x$.
  The adiabatic Hamiltonian takes form  of (\ref{1})-(\ref{1'}), where $x$ is the coordinate along
  the wire, $p$ is the conjugate momentum;  the symmetrization reestablishes the hermicity.

The other factors of appearance of SO interaction in the form
(\ref{1'}) are curvature-induced and torsion-induced SO
interactions \cite{we1,we2}. In the particular case of a curved
wire with axially symmetrical cross-section we have
\begin{equation}\label{curv}
    {\bf a} = -\alpha A_{11}\kappa {\bf b},
\end{equation}
 where ${\bf b}(x)$ denotes the binormal to the wire,
 $\kappa(x)$ is the curvature, $\alpha$ is
the effective SO coupling constant of bulk crystal \cite{we1},
$A_{11}=\langle(1+2q_1\partial_1)(\partial_1^2+\partial_2^2))\rangle_0$
is the matrix element on the transversal wave function of the
lowest subband of the wire. The quantity $A_{11}$ has order of the
energy of quantization in the wire.

\section*{Unitary transform}
 The SO Hamiltonian
(\ref{1'}) is the most general local expression which has the
first order in the SO constant and linear in $p$.  The other form
of the Hamiltonian (\ref{1}) is
\begin{equation}\label{5}
  {\cal H} =\frac{m}{2}v^2-\frac{a^2}{2}+V(x),
\end{equation}
where the velocity operator is $v=p/m+({\bf
a}(x)\mbox{\boldmath{$\sigma$}})$. We shall demonstrate, that the
Hamiltonian (\ref{5}) can be unitary transformed to the form with
no Pauli matrices. Let us consider an equation
\begin{equation}\label{2}
    (-i\frac{\partial}{\partial x}+
    ({\bf a}(x)\mbox{\boldmath{$\sigma$}}))U(x)=0
\end{equation}
for an operator $U(x)$ which explicitly depends on the coordinate
$x$. The solution of (\ref{2}) is
\begin{equation}\label{3}
  U(x)=1+\sum_{n=1} (-i)^n\int_0^x
  dx_1...\int_0^{x_{n-1}}dx_n  ~({\bf a}(x_1)\mbox{\boldmath{$\sigma$}})...
  ({\bf a}(x_n)\mbox{\boldmath{$\sigma$}}).
\end{equation}
 The expression (\ref{3}) can be
rewritten as x-ordered exponent (similar to t-ordering with
difference that the ordering should be done in $x$-space):
\begin{eqnarray} \nonumber
  U(x)=T_x(\exp(-i\int_0^x dx({\bf
 a}(x)\mbox{\boldmath{$\sigma$}}))))\equiv \\\label{10} \sum_n (-i)^n\frac{1}{n!}\int_0^x
  dx_1...\int_0^xdx_nT_x(({\bf a}(x_1)\mbox{\boldmath{$\sigma$}}))...
  ({\bf a}(x_n)\mbox{\boldmath{$\sigma$}}))
  ).
\end{eqnarray}
The operation $T_x$ means that all operators should be placed in
the decreasing order of $x_k$. The inverse operator $U(x)^{-1}$ is
determined by the ordering in the inverse order $T_x^- $:
\begin{equation}\label{11}
  U^{-1}(x)=T^-_x(\exp(+i\int_0^x dx({\bf
  a}(x)\mbox{\boldmath{$\sigma$}}))). \end{equation}
  The operator $U(x)$ is unitary: $U^+U=1$; one can treat  $U(x)$ as a spacial evolution operator.
  It  can be expanded on the $2\times 2$ matrix basis:
$U=(1+i({\bf d}\mbox{\boldmath{$\sigma$}}))(1+d^2)^{-1/2}$, where
the real vector ${\bf d}$ satisfies an equation
\begin{eqnarray}\label{200}
\frac{\partial \bf d}{\partial x}+{\bf a}+({\bf ad}){\bf d}-[{\bf
a d}]=0.
\end{eqnarray}

By means of the operator $U$ the wave function transforms as
$\psi(x)=U(x)\phi(x)$. The identities
$vU(x)\phi(x)=U(x)(p/m)\phi(x)$ and  $U(x)V(x)=V(x)U(x)$ are
valid, that yields the transformation rules $ U^+(x)v U(x)=  p/m$
and $U^+(x)V(x)U(x)=V(x)$. The transformed spin operator
$\sig(x)=U^+\sig U$ obeys the equation $\partial \sig(x)/\partial
x=-2[{\bf a}\sig(x)]$ and has the explicit form
$\sig(x)=(\sig+{\bf d}({\bf d}\sig))/(1+d^2)$.

Using these rules we find
\begin{equation}\label{12}
 {\cal H}'= U^{-1}{\cal H}U=\frac{p^2}{2m}+V(x)-\frac{a^2(x)}{2}.
\end{equation}
Thus, the transformation excludes the spin from the
Schr\"{o}dinger equation. The Hamiltonian (\ref{12}) immediately
yields the spin degeneracy of electron states, unless the boundary
conditions depend on spin explicitly.
 In particular, if the simple-connected wire is infinite in both direction
 and the states are localized, the boundary
conditions $\psi\to 0$ yield $\phi\to 0$. This means double spin
degeneracy (Kramers degeneracy). The delocalized states remain
double-degenerate also.
\section*{Responses}
 The unitary transformation of the Hamiltonian to the form (\ref{12}) has strong impact on different
response functions.  For example, consider linear responses of
electric current $J=\sigma E$, spin polarization $ S_i=\langle
\sigma_i \rangle/2=\gamma_iE$    and spin current $
J^S_i=\langle\{ \sigma_i, v\} \rangle/2=\sigma^S_iE$. The electric
field (tangent component) is assumed to be  constant along the
wire. These linear responses are expressed by  the Kubo formula
via the velocity or velocity-spin correlators
\begin{equation}\label{kub}
   \frac{e}{L} \int_{-\infty}^\infty\int_{-\infty}^\infty d\epsilon d\epsilon' \frac{f(\epsilon')-f(\epsilon)}
   {\epsilon'-\epsilon}\frac{i}{\epsilon'-\epsilon+i\delta}
   \mbox{Tr}\left(\delta(\epsilon-{\cal  H})v \delta(\epsilon'-{\cal  H}){\cal
   A}_i\right),
\end{equation}
where  in the case of conductivity  ${\cal A}_i$ stands for the
velocity operator $v$, for  the  spin orientation  and spin
current ${\cal A}_i=\sigma_i/2$ and  ${\cal A}_i=\{v,
\sigma_i\}/2$, respectively, $f(\epsilon)$ is the Fermi function,
$L$ is the length of the system.

More general expressions for responses in arbitrary order on the
electric field  are determined by the velocity correlators
\begin{equation}\label{109}
    \mbox{Tr}(v \delta(\epsilon_1-{\cal  H}) v\delta(\epsilon_2-{\cal  H})...v\delta(\epsilon_3-{\cal  H}) )
\end{equation}
or spin-velocity correlators
\begin{equation}\label{110}
    \mbox{Tr}(v \delta(\epsilon_1-{\cal  H}) v\delta(\epsilon_2-{\cal  H})...\sigma_i\delta(\epsilon_3-{\cal  H})
    ).
\end{equation}
Instead of the spin operator  one can write the spin current
operator  $\{\sigma_i, v\}/2$.

 Let us  unitary transform operators
inside  the trace operation using transformation $A\to U^{-1}AU$.
After the transformation the expression under $\mbox{Tr}$ in
(\ref{109}) becomes unit in the spin space.  The expression
(\ref{109}) reduces to
  \begin{equation}\label{109'}
    \mbox{Tr}((p/m)\delta(\epsilon_1-{\cal  H}')
    (p/m)\delta(\epsilon_2-{\cal  H}')...(p/m)
    \delta(\epsilon_3-{\cal  H}'))
\end{equation}
and (\ref{110}) goes to
  \begin{equation}\label{110'}
    \mbox{Tr}((p/m)\delta(\epsilon_1-{\cal  H}')
    (p/m)\delta(\epsilon_2-{\cal  H}')...\sigma_i(x)
    \delta(\epsilon_3-{\cal  H}'))=0.
\end{equation}
As a result of (\ref{109'}), the conductivity of the system with
SO interaction converts to that of the system with no SO
interaction. The eq.(\ref{110'}) follows from the identity
$\mbox{Tr}_\sigma(\sig(x))\equiv 0$, where $\mbox{Tr}_\sigma$
denotes the trace
  in the spin space. It proves  that both coefficients of
  spin polarization $ \gamma_i$  and spin current $\sigma^S_i$
  vanish. Similar conclusions can be done with respect to
  electrical responses of higher orders ({\it e.g.}, the photogalvanic
  effect) which are not subjected to SO interaction and spin
  responses  on the electric field  (e.g., stationary spin orientation
  by alternating electric field) which vanish.

  Note, that for
  proof of (\ref{110'})  it is essential the presence of {\it the
  only}    spin operator under the trace; the similar correlators,
  containing two or more spin operators do not vanish.
   Note also, that the proof can be reformulated in the terms of the wave function.
In fact, the wave function can be decomposed to the product of
spinor function $\chi(x)$, obeying the equation  $v\chi(x)=0$ and
scalar function $\phi(x)$ obeying the Schr\"{o}dinger equation
with the Hamiltonian (\ref{5}).  The separation of variables can
be done for the Green functions: they decay on a product of the
coordinate Green function $G_E(x,x')$ of the Hamiltonian (\ref{5})
and spin functions $U(x)U^+(x')$.

\section*{Possible generalizations}
In this section we consider possible generalizations of the
Hamiltonian (\ref{5}) which conserve the main conclusions. First,
we can include the electric field into the potential $V(x)$, hence
all conclusions remain valid in presence of it in any order of
magnitude.

Second, we can consider the potential as periodic (or containing
periodic part together with random one). Such potential without
the SO interaction forms the energy bands $\epsilon(p)$, where $p$
is now quasimomentum. The operator $p/m$ in SO part goes to
$\partial\epsilon/\partial p$.  Hence the resulting new
Hamiltonian can be also converted to the form with no spin
operators.

Third, the spin can be treated as a quantum number, counting any
pair-degenerate levels. For example, they can be subbands,
originated from two equivalent valleys of bulk semiconductor. The
Hamiltonian (\ref{1}) in that case refers to the system with
valley degeneracy without spin. According to the found transform,
the valley degeneracy will not be lifted.

Fourth, we can include spin-independent e-e interaction. As such
Hamiltonian does not touch the spin, the transformation can be
done also.

\section*{What limits the spin elimination?}
From said above one can conclude that there is no spin-orbit
interaction in 1D system. In fact, this is not the case. The spin
does not commute with the Hamiltonian (\ref{5}). Hence, an
electron with a preset spin, once injected into the wire, will
change the spin during propagation along the wire.

In particular, this manifests itself in the systems with magnetic
spin injectors/spin-selective drains \cite{schmidt}, where the
boundary conditions break the form of the Hamiltonian (\ref{5}).
(In the magnetic injector one should supplement the Hamiltonian
with the exchange term like $J\sig \bXi$, where $\bXi$ is the mean
spin density in the contact, $J$ is the exchange constant).
Conductance of a finite wire with spin-selective source and drain
should be sensitive to the spin evolution caused by the SO
interaction. Thus, the total system does not obey the conditions
of the proof.

The same is valid for cyclic systems, {\it e.g.} a ring. The
periodic boundary condition in the ring  of length $L$,
$\psi(L)=\psi(0)$ converts into the equation
$U(L)\phi(L)=\phi(0)$, containing the spin via the operator $U$.
Hence, the spin operator, being eliminated from the
Schr\"{o}dinger equation, appears in the boundary conditions that
produces the spin splitting of levels.

We have neglected the Zeeman term in the Hamiltonian, direct
interaction of spin with the magnetic field. This term actually
leads to the spin-flip transitions caused by the alternating
magnetic field and  other effects. Due to relativistic smallness
they are weak. An example of such effect is examined below.
\section*{Example: EPR-induced photogalvanic effect in spiral quantum wire}
 We consider here a spiral quantum wire with
circular cross-section.
\begin{figure}[ht]
\begin{center}
\centerline{\input epsf \epsfysize=8 cm \epsfbox{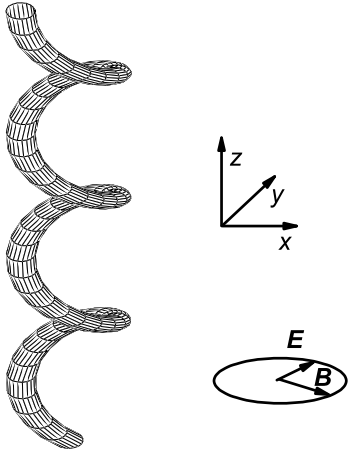}}
\end{center}
\caption{Spiral quantum wire. } \end{figure} In this system the
alternating electromagnetic field can cause the steady electric
current \cite{kib,spir}. We have previously studied  the system
neglecting the SO interaction. With taking into account  SO
interaction  the possibility of resonant current caused by
spin-flip processes arises. In accordance with said above, the
{\it electric} component of field can not induce  such current.
Hence the direct interaction of spin with magnetic field
(EPR-resonance) should be taken into account. The equation of
central line of helical wire is
\begin{equation}\label{sp}
  {\bf r}=(R\cos{(kq)},R\sin{(kq)},\eta q),
\end{equation}
where  $R$ is the radius of the helix, $q$ is the coordinate
(length) along the helix, the pitch of the helix is $2\pi
\sqrt{1/k^2-R^2}$. The sign of $k$ determines the helix direction
$\xi=\pm 1$.
 The  spiral symmetry of the wire with respect to translations along the
 wire
($q\to q+\Delta$) helps to find exact electron states.
  The adiabatic 1D Hamiltonian reads \cite{we2}
\begin{equation}\label{sp2}
  \frac{1}{2m}(p+\frac{e}{c}{\bf At})^2-\alpha A_{11}\kappa \{({\bf b}\sig ),
  (p+\frac{e}{c}{\bf At})\}
  +V(q)+\frac{1}{2}g\mu_B\sig{\bf B},
\end{equation}
where $p=-i\partial_q$,~~ ${\bf
t}(q)=(-kR\sin{(kq)},kR\cos{(kq)},\eta)$ is the tangent ort to the
wire,  ${\bf b}(q)=(\eta \sin{(kq)},\eta\cos{(kq)},kR)$ is the
binormal, ${\bf A}(t)$ is the vector-potential of electromagnetic
wave polarized in $x,y$ plane; the last (Zeemann) term describes
interaction of spin with alternating magnetic field $${\bf
B}(t)=({\bf B_0}\exp{(-i\omega t)}+c.c.)/2.$$ Without the Zeemann
term the spin can be excluded, as mentioned above and the problem
is reduced to the spinless one \cite{spir}.  The Zeemann term
results in the photogalvanic effect caused by transitions between
spin-splitted subbands.

 Let us consider the  magnetic field polarized
 in the plane $(x,y)$. The wire symmetry imposes the current
 phenomenology of the form $J_{PG}\propto \xi[{\bf B}_0,{\bf B}_0^*]_z $.
 The contribution to the stationary current due to interaction of electron spin
 with magnetic field is given by the quadratic Kubo-type formula:

 \begin{eqnarray} \nonumber
    J_{PG}=
    \frac{e}{8L}g^2\mu_B^2 ~\mbox{Re}\Bigg\{B_{0i}B_{0j}^* \int d\epsilon
     d\epsilon' d\epsilon''
  \frac{f(\epsilon')-f(\epsilon'')} {\delta+i(\omega+\epsilon'-\epsilon'')}
    \\  \label{117} \times\Big[\frac{C_{ij}(\epsilon,\epsilon',\epsilon'')}{2\delta+i(\epsilon-\epsilon'')}-
    \frac{C_{ji}(\epsilon',\epsilon'',\epsilon)}{2\delta-i(\epsilon-\epsilon')}\Big]\Bigg\},
       \end{eqnarray}
where
$C_{ij}(\epsilon,\epsilon',\epsilon'')=\mbox{Tr}(v\delta(\epsilon-{\cal
H}))\sigma_i\delta (\epsilon'-{\cal H}))
\sigma_j\delta(\epsilon''-{\cal H}))$ is the velocity-spin-spin
correlator, ${\cal H}$ is the Hamiltonian (20) in the absence of
external field (${\bf A}=0, {\bf B}=0$).
 We shall neglect complications caused by the localization of electron
 states in 1D system and emulate  the impurity scattering  by
 the switching-on field:
 the rate of the field $\delta=1/2\tau$ replaces the  reciprocal
 relaxation time $1/\tau$.

The resulting current is\begin{equation}\label{jfg}
    J_{PG} =\frac{1}{8}e\tau (g\mu_B)^2 \xi~\mbox{Im}[{\bf B}_0,{\bf
    B}_0^*]_z
    \left[
    f(\frac{2m\omega-C^2}{2C})-f(\frac{2m\omega+C^2}{2C})\right],
\end{equation}
where\begin{equation}\label{105}
    C=2m\alpha A_{11}\frac{1}{R} R^2|k|^3. \end{equation}
     The current
    exists in a narrow window of frequencies corresponding to the permitted
    spin-flip transitions.  When SO interaction is switched off
    the width of window (but not the current magnitude) shrinks.

    Thus, the direct interaction of the spin with the magnetic field of the wave results in
    the spin-guided translational effect.

The EPR-induced photogalvanic effect should be compared with the
photogalvanic effect caused by the action of electrical field on
the translational motion of electron \cite{spir}; the latter
exists in the absence of SO interaction. For a running
electromagnetic wave both effects add together, for a standing
wave (e.g., in resonator) they can be observed separately if to
place the wire in loop or node of corresponding fields. Besides,
they have different  frequency dependencies.

In conclusion, we have found that in 1D systems different response
function, which does not include the spin degree of freedom are
not influenced by spin-orbit interaction. The responses connecting
the spin and translational degrees of freedom are nonexistent
unless the direct magnetic-field spin-flip processes are taken
into account. On the contrary, the inclusion of such interaction
leads to the magnetic-field-induced resonant steady current.

In contrast to 2D systems, where SO interaction plays
determinative role for phenomena involving charge transfer and
spin, in 1D systems the influence of SO interaction is suppressed.
The transition from 2D to 1D due to lateral quantization results
in the sequential decrease of SO-induced effects.

 \section*{Acknowledgements} The
authors are grateful to A.V.Chaplik and E.G. Batyev for useful
discussions. The work was supported by grants of RFBR No's
00-02-16377 and 02-02-16398, Program for support of scientific
schools of Russian Federation No 593.2003.2 and INTAS No
03-51-6453.

\end{document}